\documentclass[twocolum,aps]{revtex4}
\usepackage{graphicx}
 
\begin{document}
\preprint{}

\title{Superfluid-insulator transition of the
Josephson junction array model with commensurate frustration}

\author{Hunpyo~Lee and Min-Chul~Cha}
\affiliation{
Department of Physics, Hanyang University, Ansan 425-791, Korea
}
\date{\today}

\begin{abstract}
We have studied the rationally frustrated Josephson-junction array model in the square
lattice through Monte Carlo simulations of $(2+1)$D $XY$-model.
For frustration $f=1/4$, the model at zero temperature shows
a continuous superfluid-insulator transition.
From the measurement of the correlation function and the superfluid stiffness,
we obtain the dynamical critical exponent $z=1.0$ and the correlation
length critical exponent $\nu=0.4 \pm 0.05$. While the dynamical critical exponent is
the same as that for cases $f=0$, $1/2$, and $1/3$,
the correlation length critical exponent is surprisingly quite different.
When $f=1/5$, we have the nature of a first-order transition.
\end{abstract}

\maketitle
The Josephson-junction array model has attracted a great deal of attention as
a prototypical model to describe the zero-temperature quantum phase transition\cite{Sondhi97}
in two dimensions. The competition of the charging energy and the Josephson
coupling energy yields the transition between superfluid and an insulator.
Recent development of the nanotechnology makes it possible to tune the
transition by changing the capacitance and the Josephson coupling of the arrays
in experiments\cite{Zant96}. The transition is also tuned by applying a magnetic field
perpendicular to the array\cite{Zant92}. The magnetic field generates frustration
in the phase coupling between the junctions.
The feature of this model describing the quantum mechanical phase transition has been widely
adopted to explain the superconductor-insulator transition in thin films\cite{Markovic99}.

The zero-temperature phase transition of the unfrustrated ($f=0$)
Josephson-junction array model belongs\cite{Doniach81,Fisher89,Cha91} to the
3D $XY$-model universality class with the dynamical critical exponent $z=1$ and
the correlation length critical exponent $\nu=0.67$\cite{Li89}.
The transitions of the model with frustration $f=1/2$ and $f=1/3$ have the
critical exponents numerically very close to those for $f=0$\cite{Cha94}. It is useful to
treat the model at $f=1/q$ in terms of $q$ coupled unfrustrated $XY$-rotors (i.e.,
order parameters) for small $q$\cite{Choi85b,Granato90}. In terms of vortices, at $f=1/2$, for example,
vortices form two different stable structures and the transition is mostly derived
by excitations with respect to them. However, the couplings between 
rotors is believed to be always relevant.  As a result, one may
expect different critical exponents for different value of $q$.

Furthermore, when frustration is very small so that the number of order parameters is large,
the first-order nature of the transition will emerge. 
This situation is analogous to the Potts model which shows
the crossover from the continuous transition to the first-order
transition as the number of the flavors increases\cite{Lee90}.
This conjecture is supported by the numerical simulation of the 3D
$XY$-model with frustration $f=1/6$ in a triangular lattice\cite{Hetzel92},
showing a first-order transition. It is also supported experimentally
by the first-order melting of dilute vortex-lines in high-${\rm T_c}$
superconductors\cite{Safar92}, whose underlying physics
can be captures by the 3D $XY$-model with small frustration. 
It is, therefore, interesting to investigate the
nature of the transition and to measure the associated critical exponents,
if it is a continuous phase transition,
in the weakly frustrated Josephson-junction array model.

In this work, we study the zero-temperature phase transition of the
Josephson-junction array model in a square lattice when the frustration is $f=1/4$
and $f=1/5$.
The model can be mapped into $(2+1)$D classical $XY$-model with frustration on the $xy$-plane.
Through Monte Carlo simulations, we find that when $f=1/4$ the superfluid-insulator transition
of the model is continuous with the dynamical critical exponent $z=1.0$
and the correlation length critical exponent $\nu=0.4 \pm 0.05$.
While the dynamical exponent is numerically
the same value as found for cases $f=0$, $1/2$, and $1/3$,
the correlation length critical exponent is surprisingly different.
When $f=1/5$, the probability distribution near the transition strongly suggests that
the nature of the transition is of first-order.

The Josephson junction array model is represented by the Hamiltonian
\begin{eqnarray}
H = \sum_i E_c n_i^2 - \sum_{\langle ij \rangle} E_J \cos(\theta_i - \theta_j - A_{ij})
\end{eqnarray}
where $E_c$ and $E_J$ denote the charging energy and the
Josephson coupling energy respectively,
$\theta_i$ the phase of the superconducting order parameter at site $i$,
$n_i = (1/i)(\partial / {\partial \theta_i})$ the number of particles at site $i$,
$A_{ij}$ the phase shift due to the magnetic field,
and $\langle ij \rangle$ pairs of nearest neighboring sites.
Frustration $f$ is embedded in $A_{ij}$ whose plaquette sum is constant over the
whole lattice, $\sum A_{ij} =2\pi f$.
It is convenient to use the notation $i=(i_x,i_y)$ to represent
the position of the $i$-th site.
Using the Landau gauge, one may have
$$ A_{ij}= \cases{
\pm 2\pi f i_x  &  if  $j=(i_x,i_y \pm1)$; \cr
0 &  otherwise. \cr}
$$

When $E_c \gg E_J$, the system has the insulating ground state,
while in the opposite limit it has the superfluid ground state.
One may treat this transition more conveniently using the equivalent  classical
effective action\cite{Cha91}
\begin{eqnarray}
S[\theta]=- K \sum_{\tilde i} [\cos (\theta_{\tilde i}-\theta_{\tilde i +\delta_x})+
\cos (\theta_{\tilde i}-\theta_{\tilde i +\delta_y}-2\pi f \tilde i_x)+
\frac{1}{2}\cos (\theta_{\tilde i}-\theta_{\tilde i +\delta_\tau} )]\ ,
\end{eqnarray}
where ${\tilde i}=(\tilde i_x,\tilde i_y,\tilde i_\tau)$ denotes the discrete lattice sites of the (2+1)D lattice
($\tilde i +\delta_x=(\tilde i_x+1,\tilde i_y,\tilde i_\tau)$, etc. ),
and $K$ the tuning parameter which effectively changes the ratio of
the Josephson coupling energy to the charging energy. We multiply the coupling in the $\tau$-direction
by a constant factor $\frac{1}{2}$ to make the superfluid stiffness and the compressibility,
which are discussed below, comparable.
Here the partition function is given by 
\begin{eqnarray}
Z={\rm Tr}_\theta e^{-S[\theta]} \ .
\end{eqnarray}

\begin{figure}
\includegraphics*[width= 7 cm]{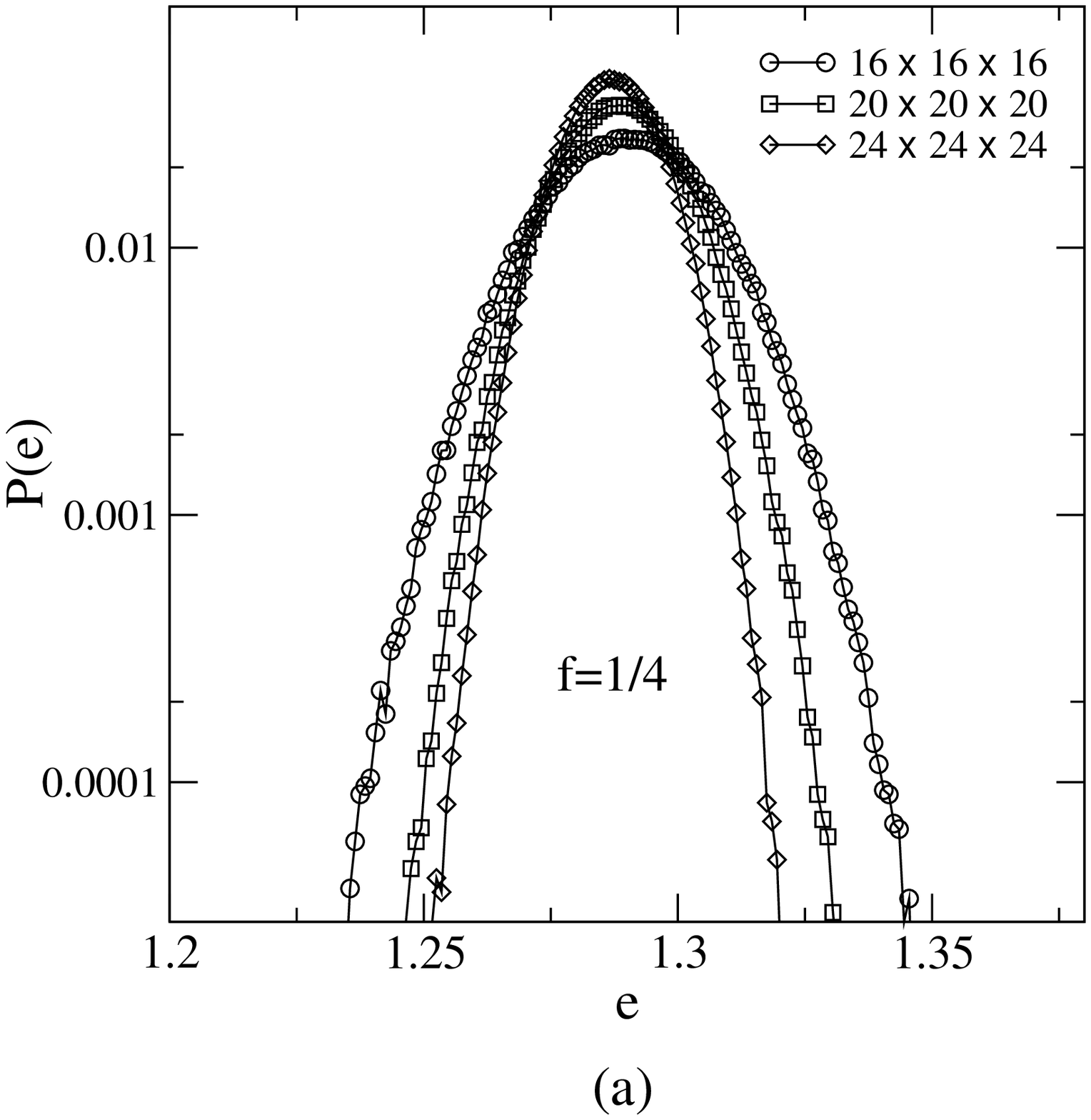}
\includegraphics*[width= 7 cm]{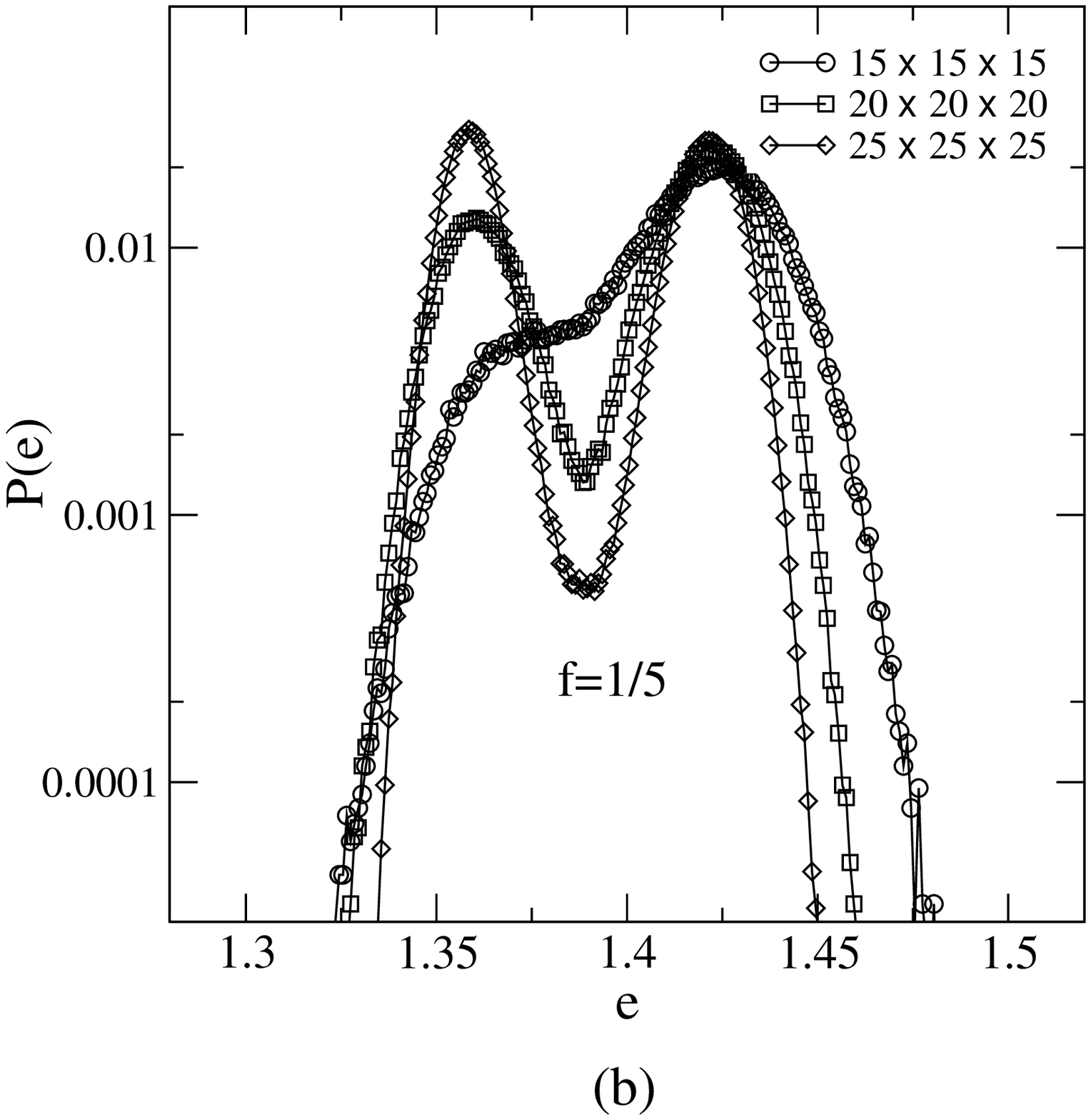}
\caption{\label{fig:Fig1}
The probability distribution as a function of the bond energy of the classical action
(a) at $f=1/4$ and (b) at $f=1/5$. The double maxima feature, the signal of the first-order transition,
appears clearly in (b), but it is absent in (a).}
\end{figure}

We simulate the model given by the classical action in the systems
whose sizes are represented by $L \times L \times L_{\tau}$. 
The first issue we are interested is the nature of the transition.
In terms of vortices, the transition is described as the vortex-line melting.
We measure the probability distribution near the expected critical point
with respect to the bond energy, $e$, of the classical action, defined by
\begin{eqnarray}
e={1 \over {L^2 L_\tau}} \sum_{\tilde i} [\cos (\theta_{\tilde i}-\theta_{\tilde i +\delta_x})+
\cos (\theta_{\tilde i}-\theta_{\tilde i +\delta_y}-2\pi f \tilde i_x)+
\frac{1}{2}\cos (\theta_{\tilde i}-\theta_{\tilde i +\delta_\tau} )]\ ,
\end{eqnarray}
Figure \ref{fig:Fig1} shows the probability distribution when $f=1/4$ and $f=1/5$.
Double maxima, separated by the distance of the latent heat and
corresponding to the two coexisting phases in a first-order transition\cite{Lee90},
do not appear when $f=1/4$. This strongly suggests that we have the continuous
nature of the transition in this case.
When $f=1/5$, however, the double maxima in the probability distribution do
appear, indicating that the nature of the transition is of first-order.

\begin{figure}
\includegraphics*[width= 7 cm]{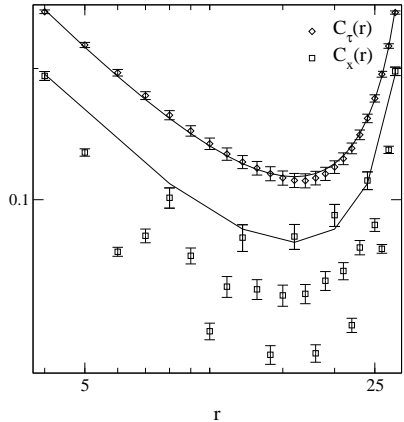}
\caption{\label{fig:Fig2}
Log-log representation of the correlation function along a spatial direction, $C_x(r)$, 
and the imaginary temporal direction, $C_\tau (r)$,
measured in the lattice whose size is $32 \times 32 \times 32$.
The solid lines represent the fitting curves.
Because of the frustration the spatial spin-spin correlation oscillates
as a function of the distance between spins. When $f=1/4$ the periodicity is 4,
as shown in the figure. This means that only those spins separated by multiples 
of 4 lattice constants are correlated each other in phase. 
The fitting curve of $C_x(r)$ represents the correlation of those spins.}
\end{figure}

For the continuous phase transition at $f=1/4$, we measure the critical exponents.
Since the transition is a quantum phase transition, the correlation length diverges
along the (imaginary) temporal direction as well as along the spatial
direction. This requires prior knowledge of the dynamical critical exponent
to determine the aspect ratio, $L_{\tau}/L^z$, of the finite size systems
in order to use the finite-size scaling analyses.
The dynamical critical exponent can be determined from the measurement of
the correlation function.
In finite size systems, the correlation function in a spatial direction, say $x$, and
in the imaginary temporal direction will follow the form 
\begin{eqnarray}
C_x (r) \sim r^{-y_x}+(L-r)^{-y_x}\ , \\
C_\tau (r) \sim r^{-y_\tau}+(L_\tau -r)^{-y_\tau} \ ,
\end{eqnarray}
respectively with $y_x=(d-2+z+\eta)$ and $y_\tau = (d-2+z+\eta)/z$.
Here $d=2$ is the spatial dimensionality.
Figure \ref{fig:Fig2} shows the correlation function measured at $K=1.296$ in the
lattice of $32 \times 32 \times 32$.
Here we obtain that $y_x \approx1.054$ and $y_\tau \approx 1.049$, implying that $z=1.0$
and $\eta \approx 0.05$.

In finite size systems the superfluid stiffness, the stiffness in a spatial direction, say $x$,
scales with respect to the diverging correlation length.
In the Landau gauge used above, the superfluid stiffness is given by
\begin{eqnarray}
\rho={1 \over {L^2 L_\tau}} \langle K \sum_{\tilde i} \cos (\theta_{\tilde i}-\theta_{\tilde i +\delta_x})
-(K \sum_{\tilde i} \sin (\theta_{\tilde i}-\theta_{\tilde i+\delta_x}))^2 \rangle \ ,
\end{eqnarray}
where $\langle \cdots \rangle$ means the average over the action of Eq.~(2).
The finite-size scaling ansatz of the superfluid stiffness can be
expressed as\cite{Cha91}
\begin{eqnarray}
\rho={1 \over L^{d+z-2}}{\tilde \rho}(L^{1/\nu}\delta, L_\tau /L^z)\ ,
\end{eqnarray}
where $\tilde \rho$ is a scaling function,
and $\delta = K-K_c$ is the tuning parameter.
The compressibility, which is the stiffness in the imaginary temporal direction,
is another quantity showing the scaling. 
The scaling behavior of compressibility will follow\cite{Cha91} 
\begin{eqnarray} 
\kappa ={1 \over L^{d-z}}{\tilde \kappa}(L^{1/\nu}\delta, L_\tau /L^z)\ , 
\label{comp_scaling} 
\end{eqnarray} 
where again ${\tilde \kappa}$ is a scaling function. 

\begin{figure}
\includegraphics*[width= 7 cm]{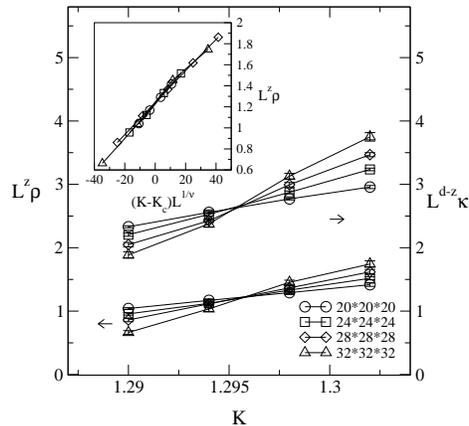}
\caption{\label{fig:Fig3}
The finite-size scaling behavior of the superfluid stiffness and the compressibility.
Curves for different sizes cross at a point, supporting that the dynamical critical exponent is $z=1.0$.
Inset: The curves collapse into a single curve with $\nu=0.4 \pm 0.05$ with respect to the scaling
variable $L^{1/\nu}(K-K_c)$.}
\end{figure}

In order to obtain the scaling behavior depending on the size of the systems,
we assume $z=1$ and take lattices of four different sizes:
i.e., $20 \times 20 \times 20, 24 \times 24 \times 24, 
28 \times 28 \times 28$, and $32 \times 32 \times 32$. 
Fig. \ref{fig:Fig3} shows the scaling behavior
of the superfluid stiffness and the compressibility.
The curves of $L^z\rho$ and $L^{d-z}\kappa$ for many different $L$
cross at $K_c=1.2958 \pm 0.001$,
implying that the chosen value of the dynamical exponent is correct.
The correlation length exponent $\nu$ is extracted by plotting
the curves with respect to $L^{1/\nu}(K-K_c)$.
We find that with $\nu=0.4 \pm 0.05$
the curves of $L^z \rho$ collapse onto a single scaling function quite well
as shown in the inset of Fig. \ref{fig:Fig3}.
The same critical exponents are obtained from the scaling behavior of the comressibility.
It is interesting to note that when $f=1/4$ the value of $\nu$ deviates from
the one for $f=0$, $f=1/2$, and $f=1/3$.

In summary, we study the superfluid-insulator phase transition of the weakly 
frustrated Josephson junction array model in two dimensions. 
Using Monte Carlo simulations, we investigate the nature of the transition and measure 
the critical exponents from the scaling behavior near the transition.
We find that when $f=1/5$ the nature of transition is of first-order.
When $f=1/4$, we have the continuous transition with the dynamical critical exponent $z=1.0$ 
and the correlation length critical exponent $\nu=0.4 \pm 0.05$.
The value of $\nu$ is different from the one for the transitions at $f=0, 1/2$ and $1/3$.
 
\begin{acknowledgments}
The authors wish to acknowledge the financial support
of Hanyang University, made in the program year of 2001, and
the grant of Korea science and engineering foundation
through the Quantum-functional Semiconductor Research Center
at Dongguk university.
\end{acknowledgments}


\begin{thebibliography}{14}
\expandafter\ifx\csname natexlab\endcsname\relax\def\natexlab#1{#1}\fi
\expandafter\ifx\csname bibnamefont\endcsname\relax
  \def\bibnamefont#1{#1}\fi
\expandafter\ifx\csname bibfnamefont\endcsname\relax
  \def\bibfnamefont#1{#1}\fi
\expandafter\ifx\csname citenamefont\endcsname\relax
  \def\citenamefont#1{#1}\fi
\expandafter\ifx\csname url\endcsname\relax
  \def\url#1{\texttt{#1}}\fi
\expandafter\ifx\csname urlprefix\endcsname\relax\def\urlprefix{URL }\fi
\providecommand{\bibinfo}[2]{#2}
\providecommand{\eprint}[2][]{\url{#2}}

\bibitem[{\citenamefont{Sondhi et~al.}(1997)\citenamefont{Sondhi, Girvin,
  Carini, and Sahar}}]{Sondhi97}
\bibinfo{author}{\bibfnamefont{S.~L.} \bibnamefont{Sondhi}},
  \bibinfo{author}{\bibfnamefont{S.~M.} \bibnamefont{Girvin}},
  \bibinfo{author}{\bibfnamefont{J.~P.} \bibnamefont{Carini}},
  \bibnamefont{and} \bibinfo{author}{\bibfnamefont{D.}~\bibnamefont{Sahar}},
  \bibinfo{journal}{Rev. Mod. Phys.} \textbf{\bibinfo{volume}{69}},
  \bibinfo{pages}{315} (\bibinfo{year}{1997}).

\bibitem[{\citenamefont{van~der Zant et~al.}(1996)\citenamefont{van~der Zant,
  Elion, Geerligs, and Mooij}}]{Zant96}
\bibinfo{author}{\bibfnamefont{H.~S.~J.} \bibnamefont{van~der Zant}},
  \bibinfo{author}{\bibfnamefont{W.~J.} \bibnamefont{Elion}},
  \bibinfo{author}{\bibfnamefont{L.~J.} \bibnamefont{Geerligs}},
  \bibnamefont{and} \bibinfo{author}{\bibfnamefont{J.~E.} \bibnamefont{Mooij}},
  \bibinfo{journal}{Phys. Rev. B} \textbf{\bibinfo{volume}{54}},
  \bibinfo{pages}{10081} (\bibinfo{year}{1996}).

\bibitem[{\citenamefont{van~der Zant et~al.}(1992)\citenamefont{van~der Zant,
  Fritschy, Elion, Geerligs, and Mooij}}]{Zant92}
\bibinfo{author}{\bibfnamefont{H.~S.~J.} \bibnamefont{van~der Zant}},
  \bibinfo{author}{\bibfnamefont{F.~C.} \bibnamefont{Fritschy}},
  \bibinfo{author}{\bibfnamefont{W.~J.} \bibnamefont{Elion}},
  \bibinfo{author}{\bibfnamefont{L.~J.} \bibnamefont{Geerligs}},
  \bibnamefont{and} \bibinfo{author}{\bibfnamefont{J.~E.} \bibnamefont{Mooij}},
  \bibinfo{journal}{Phys. Rev. Lett.} \textbf{\bibinfo{volume}{69}},
  \bibinfo{pages}{2971} (\bibinfo{year}{1992}).

\bibitem[{\citenamefont{Markovi\'c et~al.}(1999)\citenamefont{Markovi\'c,
  Christiansen, Mack, Huber, and Goldman}}]{Markovic99}
\bibinfo{author}{\bibfnamefont{N.}~\bibnamefont{Markovi\'c}},
  \bibinfo{author}{\bibfnamefont{C.}~\bibnamefont{Christiansen}},
  \bibinfo{author}{\bibfnamefont{A.~M.} \bibnamefont{Mack}},
  \bibinfo{author}{\bibfnamefont{W.~H.} \bibnamefont{Huber}}, \bibnamefont{and}
  \bibinfo{author}{\bibfnamefont{A.~M.} \bibnamefont{Goldman}},
  \bibinfo{journal}{Phys. Rev. B} \textbf{\bibinfo{volume}{60}},
  \bibinfo{pages}{4320} (\bibinfo{year}{1999}).

\bibitem[{\citenamefont{Doniach}(1981)}]{Doniach81}
\bibinfo{author}{\bibfnamefont{S.}~\bibnamefont{Doniach}},
  \bibinfo{journal}{Phys.Rev. B} \textbf{\bibinfo{volume}{24}},
  \bibinfo{pages}{5063} (\bibinfo{year}{1981}).

\bibitem[{\citenamefont{Fisher et~al.}(1989)\citenamefont{Fisher, Weichman,
  Grinstein, and Fisher}}]{Fisher89}
\bibinfo{author}{\bibfnamefont{M.~P.~A.} \bibnamefont{Fisher}},
  \bibinfo{author}{\bibfnamefont{P.~B.} \bibnamefont{Weichman}},
  \bibinfo{author}{\bibfnamefont{G.}~\bibnamefont{Grinstein}},
  \bibnamefont{and} \bibinfo{author}{\bibfnamefont{D.~S.}
  \bibnamefont{Fisher}}, \bibinfo{journal}{Phys. Rev. B}
  \textbf{\bibinfo{volume}{40}}, \bibinfo{pages}{546} (\bibinfo{year}{1989}).

\bibitem[{\citenamefont{Cha et~al.}(1991)\citenamefont{Cha, Fisher, Girvin,
  Wallin, and Young}}]{Cha91}
\bibinfo{author}{\bibfnamefont{M.-C.} \bibnamefont{Cha}},
  \bibinfo{author}{\bibfnamefont{M.~P.~A.} \bibnamefont{Fisher}},
  \bibinfo{author}{\bibfnamefont{S.~M.} \bibnamefont{Girvin}},
  \bibinfo{author}{\bibfnamefont{M.}~\bibnamefont{Wallin}}, \bibnamefont{and}
  \bibinfo{author}{\bibfnamefont{A.~P.} \bibnamefont{Young}},
  \bibinfo{journal}{Phys. Rev. B} \textbf{\bibinfo{volume}{44}},
  \bibinfo{pages}{6883} (\bibinfo{year}{1991}).

\bibitem[{\citenamefont{Li and Teitel}(1989)}]{Li89}
\bibinfo{author}{\bibfnamefont{Y.-H.} \bibnamefont{Li}} \bibnamefont{and}
  \bibinfo{author}{\bibfnamefont{S.}~\bibnamefont{Teitel}},
  \bibinfo{journal}{Phys. Rev. B} \textbf{\bibinfo{volume}{40}},
  \bibinfo{pages}{9122} (\bibinfo{year}{1989}).

\bibitem[{\citenamefont{Cha and Girvin}(1994)}]{Cha94}
\bibinfo{author}{\bibfnamefont{M.-C.} \bibnamefont{Cha}} \bibnamefont{and}
  \bibinfo{author}{\bibfnamefont{S.~M.} \bibnamefont{Girvin}},
  \bibinfo{journal}{Phys. Rev. B} \textbf{\bibinfo{volume}{49}},
  \bibinfo{pages}{9794} (\bibinfo{year}{1994}).

\bibitem[{\citenamefont{Choi and Doniach}(1985)}]{Choi85b}
\bibinfo{author}{\bibfnamefont{M.~Y.} \bibnamefont{Choi}} \bibnamefont{and}
  \bibinfo{author}{\bibfnamefont{S.}~\bibnamefont{Doniach}},
  \bibinfo{journal}{Phys. Rev. B} \textbf{\bibinfo{volume}{31}},
  \bibinfo{pages}{4516} (\bibinfo{year}{1985}).

\bibitem[{\citenamefont{Granato and Kosterlitz}(1990)}]{Granato90}
\bibinfo{author}{\bibfnamefont{E.}~\bibnamefont{Granato}} \bibnamefont{and}
  \bibinfo{author}{\bibfnamefont{J.~M.} \bibnamefont{Kosterlitz}},
  \bibinfo{journal}{Phys. Rev. Lett.} \textbf{\bibinfo{volume}{65}},
  \bibinfo{pages}{1267} (\bibinfo{year}{1990}).

\bibitem[{\citenamefont{Lee and Kosterlitz}(1990)}]{Lee90}
\bibinfo{author}{\bibfnamefont{J.}~\bibnamefont{Lee}} \bibnamefont{and}
  \bibinfo{author}{\bibfnamefont{J.~M.} \bibnamefont{Kosterlitz}},
  \bibinfo{journal}{Phys. Rev. Lett.} \textbf{\bibinfo{volume}{65}},
  \bibinfo{pages}{137} (\bibinfo{year}{1990}).

\bibitem[{\citenamefont{Hetzel et~al.}(1992)\citenamefont{Hetzel, Sudb\o, and
  Huse}}]{Hetzel92}
\bibinfo{author}{\bibfnamefont{R.~E.} \bibnamefont{Hetzel}},
  \bibinfo{author}{\bibfnamefont{A.}~\bibnamefont{Sudb\o}}, \bibnamefont{and}
  \bibinfo{author}{\bibfnamefont{D.~A.} \bibnamefont{Huse}},
  \bibinfo{journal}{Phys. Rev. Lett.} \textbf{\bibinfo{volume}{69}},
  \bibinfo{pages}{518} (\bibinfo{year}{1992}).

\bibitem[{\citenamefont{Safar et~al.}(1992)\citenamefont{Safar, Gammel, Huse,
  Bishop, Rice, and Ginsberg}}]{Safar92}
\bibinfo{author}{\bibfnamefont{H.}~\bibnamefont{Safar}},
  \bibinfo{author}{\bibfnamefont{P.~L.} \bibnamefont{Gammel}},
  \bibinfo{author}{\bibfnamefont{D.~A.} \bibnamefont{Huse}},
  \bibinfo{author}{\bibfnamefont{D.~J.} \bibnamefont{Bishop}},
  \bibinfo{author}{\bibfnamefont{J.~P.} \bibnamefont{Rice}}, \bibnamefont{and}
  \bibinfo{author}{\bibfnamefont{D.~M.} \bibnamefont{Ginsberg}},
  \bibinfo{journal}{Phys. Rev. Lett.} \textbf{\bibinfo{volume}{69}},
  \bibinfo{pages}{824} (\bibinfo{year}{1992}).

\end{thebibliography}
\end{document}